\documentclass[reqno]{amsart}
\usepackage{graphicx,graphics,color} 
\usepackage{amsmath,amssymb,enumitem,mathtools}
\usepackage[normalem]{ulem}
\usepackage{subcaption}
\usepackage{float}
\usepackage[permil]{overpic}
\usepackage[backend=biber,style=nature,url=false,doi=false,isbn=false]{biblatex}
\usepackage{url}
\usepackage{fullpage}
\usepackage{amsaddr}

\begin{document}
\pagestyle{plain}

\title{\Large Cell Division Changes Fate Decisions \\ in a Genetic Toggle Switch}

\author{Charli Austin}
\address{School of Biological Sciences, University of Edinburgh, Edinburgh, U.K.}
\author{Nikola Popovi\'c}
\address{School of Mathematics and Maxwell Institute for Mathematical Sciences, University of Edinburgh, Edinburgh, U.K.}
\author{Ramon Grima$^\ast$}
\address{School of Biological Sciences, University of Edinburgh, Edinburgh, U.K.}
\address{$^\ast$Corresponding author; ramon.grima@ed.ac.uk}



\begin{abstract}

Gene regulatory networks govern cellular fate decisions through multistable dynamics. The genetic toggle switch is a canonical model of such behaviour; yet, the impact of cell division on its dynamics remains poorly understood. We derive analytical separatrices for a simplified Boolean toggle switch with and without division. We show that division can redirect trajectories with identical initial conditions to opposing stable states, and we define a region of disagreement where fate decisions are predicted incorrectly if division is neglected. Our results imply that division can fundamentally reshape fate boundaries in multistable regulatory networks.

\end{abstract}

\maketitle

\section*{Introduction}

Gene regulatory networks (GRNs) comprise interactions among genes, transcription factors, and other molecular regulators that govern gene expression.
These networks integrate intracellular and extracellular signals to regulate gene activity, thereby controlling a wide range of cellular processes \cite{Maizels2026}.
Of particular interest are cellular fate decisions, in which regulatory networks drive transitions between alternative stable cellular states, such as differentiation and programmed cell death \cite{Balazsi2011,Perkins2009,ham2025mapping}.

A popular and well-studied example of these GRNs is the genetic toggle switch, a synthetic bistable circuit introduced by Ref.~\cite{Gardner2000}. Toggle-switch architectures also arise naturally in a variety of biological systems, such as the Lambda phage lysis-lysogeny switch \cite{Lee2018, Atsumi2006, Oppenheim2005, Tian2004} and the PU.1--GATA-1 interaction \cite{Chickarmane2009,Duff2011}. 
The toggle switch is a mutually inhibitory network composed of two genes, each of which produces a protein that can suppress the activity of the opposing gene. It is one of the simplest circuits displaying robust bistable behaviour \cite{Gardner2000, Lipshtat2006}. Bistable systems are strongly linked to cellular decision-making, with each steady state often corresponding to a different decision or fate. The toggle switch has been linked to cellular decision-making both in its own right \cite{Strasser2012,Cinquin2005} and as a proxy in models for more complex two-decision systems, such as the epithelial-mesenchymal transition in developmental and cancer biology \cite{Biswas2022}.

While classical models of gene expression incorporate the effects of cell division via the inclusion of effective dilution in the overall degradation rate, explicit division mechanisms only remove these proteins at the end of the cycle. Thus, the respective models are different dynamically --- these differences are apparent even in models that do not include feedback regulation \cite{beentjes2020exact}. Currently, it is not clear how the explicit modelling of cell division impacts GRN dynamics; one of the few studies of this type, Ref. \cite{Bierbaum2015}, found that the region of bistability is reduced when cell division effects are fully taken into account. However, for understanding cellular decision-making, it remains unclear whether, in parameter regimes where both models with and without division are bistable, identical initial conditions result in the same stable state. In this paper, we answer this question by proposing a simplified, analytically tractable version of the toggle switch, both with and without division.

We start with a brief summary of the classical toggle switch, which is modelled via the following system of coupled ODEs; for the full reaction scheme, see Appendix A: 
\begin{align}
        \frac{\text{d}N_A}{\text{dt}} &= \frac{\alpha_A}{1 + [N_B/(\lambda_BV)]^{h_1}} - (\beta_A+\delta)N_A, \label{eqn:NA_classical}
    \\
    \frac{\text{d}N_B}{\text{dt}} &= \frac{\alpha_B}{1 + [N_A/(\lambda_AV)]^{h_2}}- (\beta_B+\delta)N_B, \label{eqn:NB_classical}
\end{align}
where $N_A$ and $N_B$ are the counts of proteins $A$ and $B$, respectively; $\alpha_A$ and $\alpha_B$ are the maximum synthesis rates; $\beta_A$ and $\beta_B$ are the active degradation rates; $\delta=\ln 2/T$ is the effective dilution rate, where $T$ is the mean cell cycle length; $\lambda_A$ and $\lambda_B$ are threshold concentrations for the repression of proteins $A$ and $B$, respectively; $h_1$ and $h_2$ are Hill coefficients associated with the cooperativity of repression; and $V$ is the cell volume.

\begin{figure*}
    \centering
    \begin{overpic}[width=0.75\linewidth]{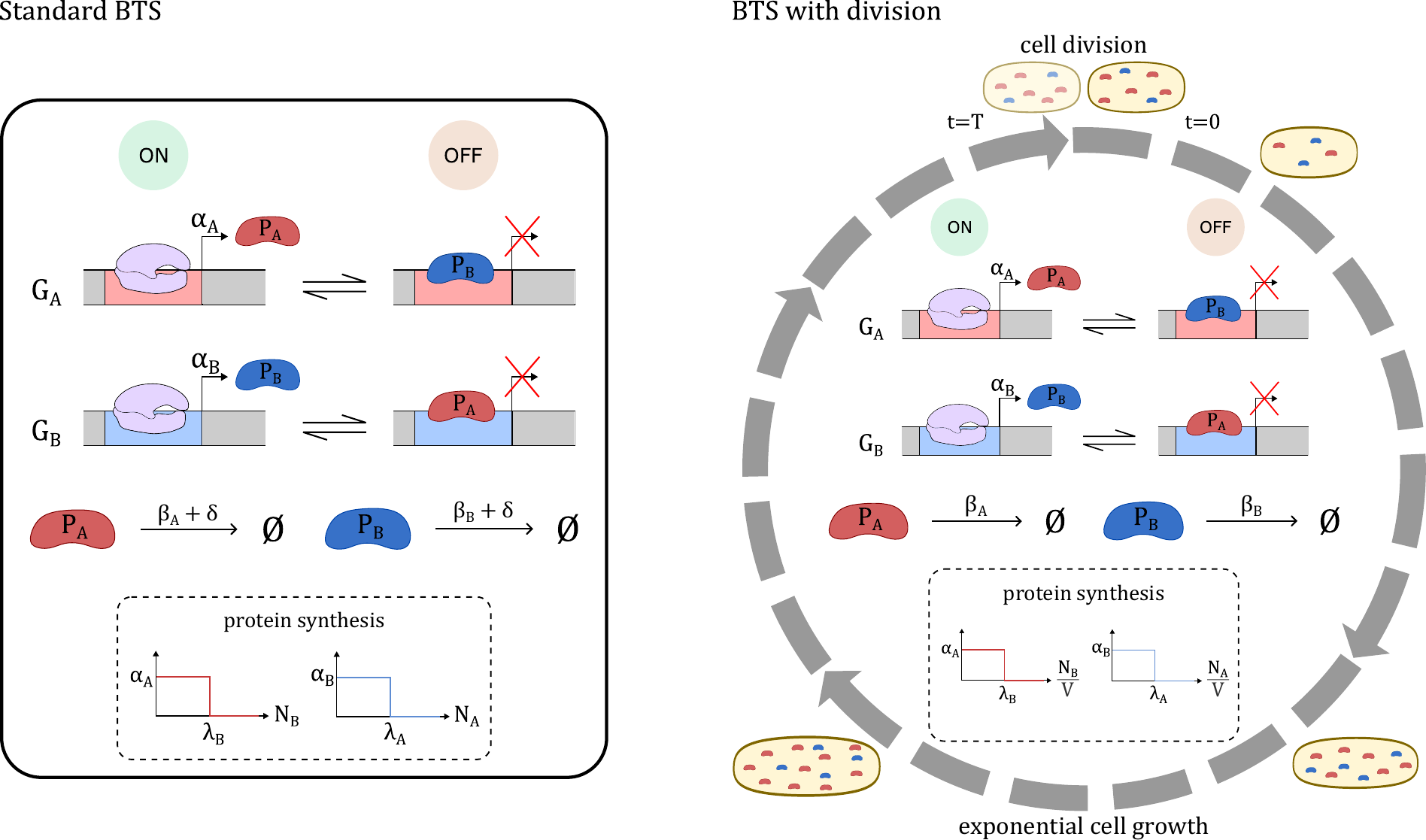}
        \put(-50,600){(a)}
        \put(460,600){(b)}
    \end{overpic}
    \caption{Illustration of two different versions of the BTS. (a) Standard BTS: the volume of the system is fixed to $V=1$ so that protein concentration and count are equivalent. 
    Mutual inhibition is determined by the protein counts reaching a fixed threshold. Proteins are removed from the system with rate $\beta_X + \delta$ to account for both active degradation with rate $\beta_X$ and effective dilution with rate $\delta$. (b) BTS with division: the cell grows exponentially, with inhibition determined by the protein concentrations reaching a fixed threshold. There is no effective dilution, as division is incorporated explicitly; every time a cell cycle is completed, the molecule count of both proteins is halved. We assume that  gene expression is buffered against changes in gene copy number and that replication therefore has no impact on the dynamics \cite{PerezOrtin2019,Voichek2016,jia2022concentration}.}
    \label{fig:BTS_cartoon}
\end{figure*}

The bistability of the system described by Eqs.~\eqref{eqn:NA_classical} and \eqref{eqn:NB_classical} depends on the chosen parameter set. In bistable regimes, two stable equilibria exist, corresponding to high expression of one protein and low expression of the other. The outcome is determined by initial conditions: the phase plane is divided into two basins of attraction by a separatrix -- the stable manifold of a saddle equilibrium -- with trajectories remaining in their basin, converging to the corresponding equilibrium \cite{Strogatz2018}. The transition between between monostability and bistability has been studied in Ref.~\cite{Li2020}. Strong nonlinearity in these models renders it difficult to study them analytically, even without division.

To address the challenge of analytical intractability, we introduce the Boolean Toggle Switch (BTS), whereby the activation function is binarised -- protein synthesis being either ON or OFF according to a fixed threshold in the opposing protein concentration. The ensuing simplification in Eqs.~\eqref{eqn:NA_classical} and \eqref{eqn:NB_classical}, whereby the Hill functions are removed, allows for more accessible analytical investigation while preserving the fundamental dynamical features of the toggle switch. In doing so, as we shall show, we can quantify the effects of cell division by classifying the differences in the separatrices for the standard BTS and the BTS with division.

\section*{Boolean Toggle Switch}

The BTS is obtained by considering Eqs. \eqref{eqn:NA_classical} and \eqref{eqn:NB_classical} in the limit of infinite Hill coefficients, allowing us to replace the Hill functions with Heaviside functions, $H(x)$, where $H(x) = 1$ for $x>0$ and $H(x) = 0$ for $x \leq 0$; see Appendix B for the full derivation.

In order to investigate the effects of cell division on the BTS, we must formulate two versions of the model -- first, the standard BTS which follows classical modelling convention, incorporating division solely through an effective dilution rate (Fig.~\ref{fig:BTS_cartoon}a); second, the BTS with division which explicitly models the division mechanism (Fig.~\ref{fig:BTS_cartoon}b).

For the standard BTS, we take the volume of the cell to be fixed, with $V=1$; hence, the corresponding governing equations read 
\begin{align}
    \frac{\text{d}N_A}{\text{dt}} &= \alpha_AH(\lambda_B - N_B) - (\beta_A + \delta)N_A,
    \label{eqn:NA_standard_BTS}
    \\
    \frac{\text{d}N_B}{\text{dt}} &= \alpha_BH(\lambda_A - N_A) - (\beta_B+\delta)N_B.
    \label{eqn:NB_standard_BTS}
\end{align}

In the BTS with division, we assume that the cell grows exponentially \cite{soifer2016single} with a fixed cycle duration $T$ and that at end of the cell cycle, the count of both proteins is halved. Specifically, the kinetic equations are now given by
\begin{align}
    \frac{\text{d}N_A}{\text{dt}} &= \alpha_AH(\lambda_B V - N_B) - \beta_AN_A,
    \label{eqn:NA_BTS}
    \\
    \frac{\text{d}N_B}{\text{dt}} &= \alpha_BH(\lambda_A V - N_A) - \beta_BN_B,
    \label{eqn:NB_BTS}
\end{align}
where $V= V_0 e^{\frac{\ln 2}{T}t}$ and $t$ is the time within the cell cycle which is reset to $0$ when division occurs at time $t=T$. The choice of $V$ ensures that $V(T) = 2V(0)$, i.e. the volume of the cell doubles over the course of one cell cycle. Furthermore, by taking $V_0 = \ln 2$, the time average of the volume over each cell cycle is equal to 1, ensuring a faithful comparison between the two  versions of the BTS. In contrast to the standard BTS, Eqs.~\eqref{eqn:NA_standard_BTS} and \eqref{eqn:NB_standard_BTS}, the dilution rate $\delta$ does not enter the removal terms, since division is explicitly modelled in the BTS with division. Note also that the protein count of only one cell is tracked, equivalent to single cell lineage tracing in experimental setups \cite{jia2021frequency,tanouchi2017long}.

Within this Boolean framework, we are able to derive analytical expressions for the two separatrices in the phase plane for both versions of the BTS (Table~\ref{tab:separatrix_expressions}).

\section*{Derivation of separatrix expressions}

From Ref.~\cite{Schwanhaeusser2011}, we find that approximately 71$\%$ of proteins in mammalian cells have a half-life that is longer than the median cell cycle length; see Fig.~\ref{fig:mu_dist}, Appendix C for the distribution. As the effect of protein degradation is then negligible within the cell cycle, we restrict to the case of zero active protein degradation, with $\beta_A=0=\beta_B$. (Derivations for non-zero active degradation are given in Appendix C.) If $\alpha_A/\delta <\lambda_A$ or $\alpha_B/\delta <\lambda_B$, the standard BTS is a monostable system and therefore admits no separatrix, see Appendix D; hence, we exclude these parameter regimes from our analysis.

We first derive the separatrix for the standard BTS. Note that for $V=1$, concentrations and protein counts are equivalent, which means that the repression threshold is also a fixed protein count.

\renewcommand{\arraystretch}{2}
\begin{table}
    \centering
        \begin{tabular}{ccc}
        \hline \hline
        Model & Segment label & Separatrix expression \\
        \hline
        Standard BTS & $S_1$ & $\frac{\alpha_A-\delta A_0}{\alpha_A-\delta\lambda_A}=\frac{\alpha_B-\delta B_0}{\alpha_B - \delta\lambda_B}$ \\
      & $S_2$ & $\frac{A_0}{\lambda_A} = \frac{B_0}{\lambda_B}$ \\
     Division BTS & $D_1$ & $\frac{A_0}{\alpha_A} + \frac{\Omega_A}{\delta} = \frac{B_0}{\alpha_B} + \frac{\Omega_B}{\delta}$ \\ 
      & $D_2$ & $\frac{A_0}{\lambda_A} = \frac{B_0}{\lambda_B}$ \\
      \hline \hline
    \end{tabular}
    
    \caption{Expressions for the separatrices for both the standard BTS and the BTS with division. The function $\Omega_X$ is defined in Eq.~\eqref{eqn:omega} in Appendix E. Note that $S_2 = D_2$.}
    \label{tab:separatrix_expressions}
\end{table}

\begin{figure*}
    \centering
    \includegraphics[width = \linewidth]{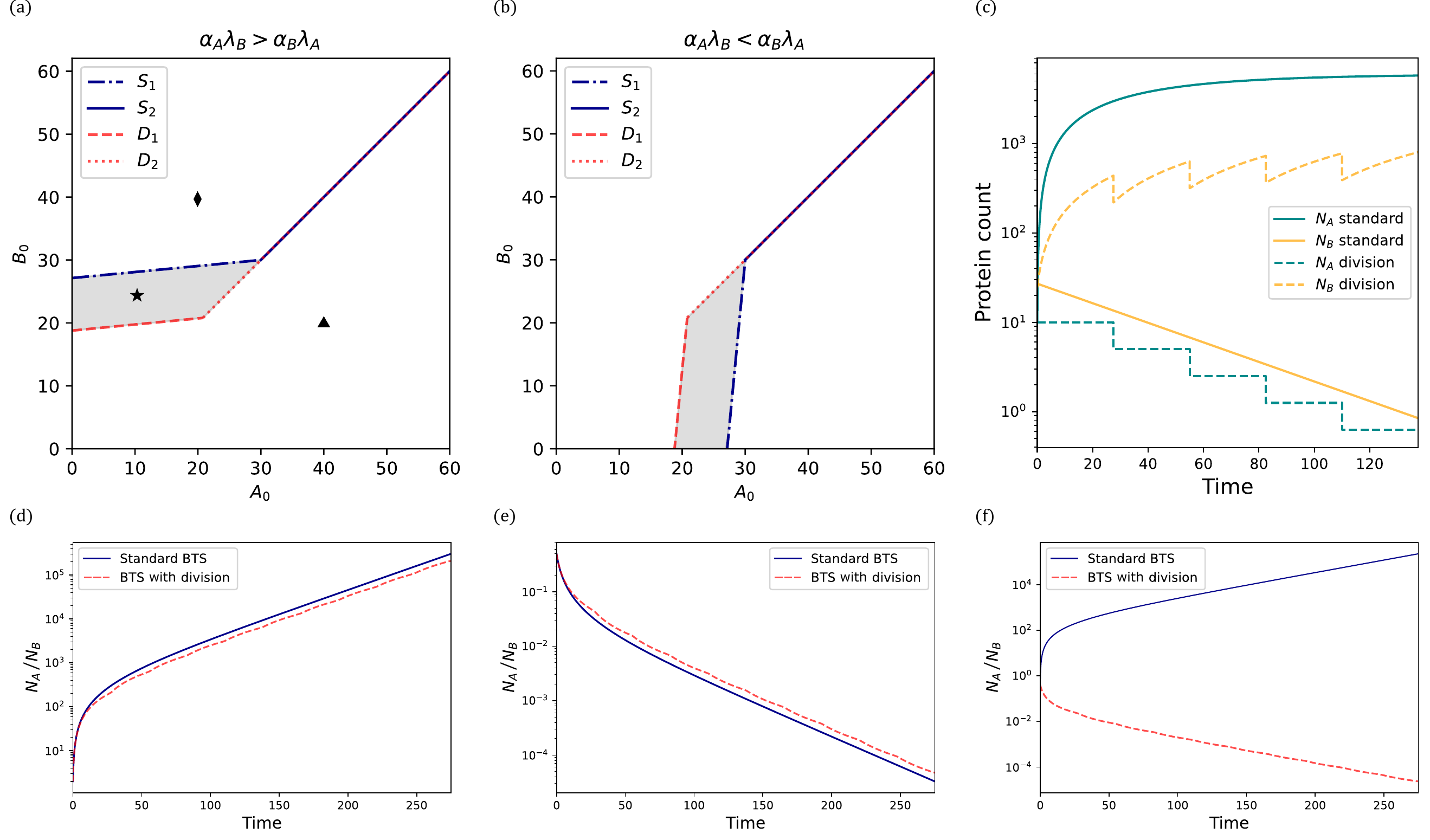}
    \caption{(a) and (b) Plots of the separatrix expressions from Table \ref{tab:separatrix_expressions} in different parameter regimes. \textcolor{black}{Shaded areas indicate the corresponding RODs.} (a) For $\alpha_A\lambda_B > \alpha_B\lambda_A$, the division separatrix lies below the standard separatrix. Solid black shapes mark representative initial conditions for each of the dynamic behaviours presented in panels (d--f). (b) For $\alpha_A\lambda_B < \alpha_B\lambda_A$, the division separatrix lies to the left of the standard separatrix. (c) An example trajectory for disagreeing dynamic behaviour, with initial conditions marked by the star in panel (a). (d--f) The change in $N_A/N_B$ over time, each with initial conditions as marked in panel (a): (d) triangle, (e) diamond, and (f) star.}
    \label{fig:fig_2}
\end{figure*}

Let $A_0$ and $B_0$ be the initial counts of proteins $A$ and $B$, respectively. The separatrix is then derived by considering four cases, based on the initial counts relative to the thresholds: (i) $A_0 < \lambda_A$, $B_0 < \lambda_B$; (ii) $A_0 \geq\lambda_A$, $B_0 \geq \lambda_B$; (iii) $A_0 < \lambda_A$, $B_0 \geq \lambda_B$; (iv) $A_0 \geq \lambda_A$, $B_0 < \lambda_B$. 

In case (i), whichever protein reaches its threshold first will be the ``winning" protein. Solving Eqs.~\eqref{eqn:NA_standard_BTS} and \eqref{eqn:NB_standard_BTS} with $\beta_A=0=\beta_B$ and rearranging for $t$, we can calculate the time at which protein $X$ ($X \in \{A,B\}$) reaches its threshold, $t_{\lambda_X}$:
\begin{equation}
    t_{\lambda_X} = \frac{1}{\delta}\ln \frac{\alpha_X - \delta X_0}{\alpha_X - \delta\lambda_X}.
\end{equation}
Note that the argument of the logarithm is always positive, since $A_0 < \lambda_A$ and $B_0 < \lambda_B$ and as bistability requires $\alpha_X/\delta > \lambda_X$.

An expression for the separatrix can then be found when both proteins reach their respective thresholds at the same time, i.e. when $t_{\lambda_A}$ = $t_{\lambda_B}$, which is given by segment $S_1$ (Table \ref{tab:separatrix_expressions}, row 1). Case (ii) gives segment $S_2$ (Table \ref{tab:separatrix_expressions}, row 2) by an analogous argument.

In cases (iii) and (iv), the protein whose count is above its threshold initially will win automatically---as there is no possibility for competition, no segment of the separatrix will lie in these regions of the phase plane.

For the BTS with division, recall that the cell grows exponentially; thus, for the concentration threshold to be constant, the protein count threshold for protein $X$, $\psi_X(t)$, must scale proportionally to the volume of the cell: $\psi_X(t) = \lambda_X V_0 e^{\delta t}$. Applying the same threshold-crossing argument as above yields segments $D_1$ and $D_2$ (rows~3 and 4 of Table~\ref{tab:separatrix_expressions}); see Appendix E for the derivation.

In Figs.~\ref{fig:fig_2}(a) and (b), we plot the resulting separatrices and note that the phase plane is divided into three regions of different dynamic behaviour: first, the region below both separatrices, where protein $A$ will ``win" in both the standard BTS and the BTS with division; second, the region above both separatrices, where protein $B$ will win in both models; finally, the enclosed Region of Disagreement (ROD) where trajectories of the two models are attracted to different steady states (Fig.~\ref{fig:fig_2}(c)). Note that the two separatrices coincide beyond $(\lambda_A,\lambda_B)$, as segments $S_2$ and $D_2$ share the same functional form, but have different domains. Figs.~\ref{fig:fig_2}(d--f) show the evolution of $N_A/N_B$ in each of these three regions.

We obtain three distinct categorisations of the ROD, depending on the parameter regime selected; see Appendix F for a full geometrical analysis:
\begin{enumerate}
    \item $\alpha_A\lambda_B > \alpha_B\lambda_A$: the ROD lies below the standard separatrix (Fig. \ref{fig:fig_2}a);
    \item $\alpha_A\lambda_B < \alpha_B\lambda_A$: the ROD lies to the left of the standard separatrix (Fig. \ref{fig:fig_2}b); and
    \item $\alpha_A\lambda_B = \alpha_B\lambda_A$: the separatrices coincide, meaning there is no ROD.
\end{enumerate}

In case 1, the area of the ROD is approximately
\begin{equation*}
    A_{\rm ROD} \approx \frac{1}{2}\lambda_A(1-V_0^2)\frac{\alpha_A\lambda_B-\alpha_B\lambda_A}{\alpha_A-\delta\lambda_A}.
\end{equation*}
 Since $V_0 = \ln 2$ and $\alpha_X/\delta > \lambda_X$ is necessary for bistability, the area is strictly positive; furthermore, $A_{\rm ROD}=0$ when $\alpha_A\lambda_B = \alpha_B\lambda_A$, as expected. Finally, the area is maximised as $\alpha_A\lambda_B \to \infty$, with the limit being $\frac{1}{2}\lambda_A\lambda_B(1-V_0^2)$. Similarly, the area for case 2 can be found by a simple switching of indices ($A\leftrightarrow B$), giving the same limit, $\frac{1}{2}\lambda_A\lambda_B(1-V_0^2)$, as $\alpha_B\lambda_A \to \infty$.

Hence, large ratios between the two protein synthesis rates or the two repression thresholds maximises the area of the ROD, increasing the probability of contradictory predictions between models. \textcolor{black}{Similar results are found numerically if, instead of the BTS, one considers the classical toggle switch (Fig.~\ref{fig:finite_hill_BTS}) with finite values for the Hill coefficients; the advantage of the BTS is its analytical tractability.}

\section*{Stochastic Simulation of the BTS}

\begin{figure}
    \centering
    \begin{overpic}[width=\linewidth]{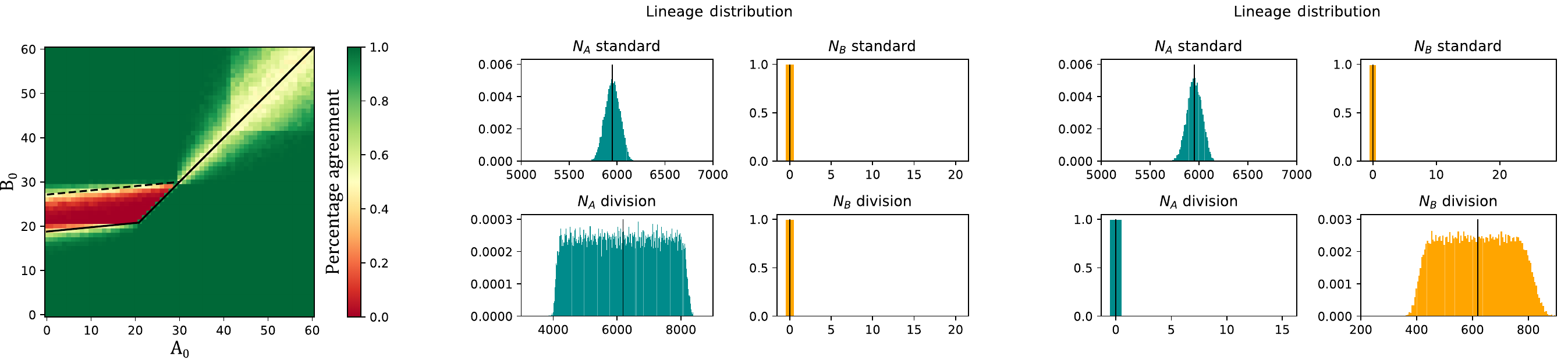}
        \put(0,212){(a)}
        \put(280,212){(b)}
        \put(650,212){(c)}
    \end{overpic}
    \caption{(a) Percentage agreement between stochastic simulations of the standard BTS and the BTS with division. Dark green denotes complete agreement of final states; dark red denotes complete disagreement; \textcolor{black}{the deterministic ROD is enclosed by dashed and solid black lines.} (b) Example lineage distributions for initial conditions in the dark green region of (a). (c) Example lineage distributions for initial conditions in the dark red region of (a). Black vertical lines indicate the deterministic steady state (standard BTS) or cyclostationary average (BTS with division).
    }
    \label{fig:stochastic_heatmaps}
\end{figure}

Since the ROD encompasses low-count initial conditions where intrinsic noise is non-negligible \cite{elowitz2002stochastic}, we performed stochastic simulations for a timer mechanism of cell division to validate our deterministic analysis. Stochastic fluctuations can cause spontaneous switching between states in bistable systems \cite{Biancalani2014,Loinger2007,PerezCarrasco2016}; as such, we expect that there will be a proportion of simulations which result in disagreement, even for initial conditions for which steady states agree deterministically. To investigate agreement between the stochastic standard BTS and the BTS with division, we perform $500$ simulations for each initial condition and record the proportion of those where steady states disagree between models, yielding an overall ``percentage of agreement"; see Appendix G for details. We find that initial conditions producing disagreement in more than 50$\%$ of simulations are well-contained within the deterministic ROD, confirming its robustness to intrinsic noise (Fig.~\ref{fig:stochastic_heatmaps}). Similar results are expected for more complex cell-division models \cite{jia2021cell}.

\section*{Summary and Outlook}

Concluding, we introduced a Boolean toggle switch model and derived closed-form separatrices with and without division. We identified a Region of Disagreement in which the two models predict opposite cellular fates from identical initial conditions. These results show that fate predictions are substantially affected by cell division and suggest that similar discrepancies may arise in more complex multistable gene regulatory networks.

\section*{Acknowledgments}
C. A acknowledges support from a PhD scholarship provided by the Engineering and Physical Sciences Research Council. R. G. acknowledges support from the Leverhulme Trust (RPG-2024-082). 

\printbibliography

@article{jia2021frequency,
  title={Frequency domain analysis of fluctuations of mRNA and protein copy numbers within a cell lineage: theory and experimental validation},
  author={Jia, Chen and Grima, Ramon},
  journal={Physical Review X},
  volume={11},
  number={2},
  pages={021032},
  year={2021},
  publisher={APS},
}

@article{ham2025mapping,
  title={Mapping, Modeling, and Reprogramming Cell-Fate Decision-Making Systems},
  author={Ham, Lucy and Woodward, Taylor E and Coomer, Megan A and Stumpf, Michael PH},
  journal={Annual Review of Biomedical Data Science},
  volume={8},
  year={2025},
  publisher={Annual Reviews}
}

@article{jia2021cell,
  title={Cell size distribution of lineage data: analytic results and parameter inference},
  author={Jia, Chen and Singh, Abhyudai and Grima, Ramon},
  journal={Iscience},
  volume={24},
  number={3},
  year={2021},
  publisher={Elsevier},
}

@article{elowitz2002stochastic,
  title={Stochastic gene expression in a single cell},
  author={Elowitz, Michael B and Levine, Arnold J and Siggia, Eric D and Swain, Peter S},
  journal={Science},
  volume={297},
  number={5584},
  pages={1183--1186},
  year={2002},
  publisher={American Association for the Advancement of Science},
}

@article{tanouchi2017long,
  title={Long-term growth data of Escherichia coli at a single-cell level},
  author={Tanouchi, Yu and Pai, Anand and Park, Heungwon and Huang, Shuqiang and Buchler, Nicolas E and You, Lingchong},
  journal={Scientific data},
  volume={4},
  number={1},
  pages={170036},
  year={2017},
  publisher={Nature Publishing Group},
}

@article{soifer2016single,
  title={Single-cell analysis of growth in budding yeast and bacteria reveals a common size regulation strategy},
  author={Soifer, Ilya and Robert, Lydia and Amir, Ariel},
  journal={Current Biology},
  volume={26},
  number={3},
  pages={356--361},
  year={2016},
  publisher={Elsevier}
}

@article{jia2022concentration,
  title={Concentration fluctuations in growing and dividing cells: Insights into the emergence of concentration homeostasis},
  author={Jia, Chen and Singh, Abhyudai and Grima, Ramon},
  journal={PLoS computational biology},
  volume={18},
  number={10},
  pages={e1010574},
  year={2022},
  publisher={Public Library of Science San Francisco, CA USA},
}

@article{beentjes2020exact,
  title={Exact solution of stochastic gene expression models with bursting, cell cycle and replication dynamics},
  author={Beentjes, Casper HL and Perez-Carrasco, Ruben and Grima, Ramon},
  journal={Physical Review E},
  volume={101},
  number={3},
  pages={032403},
  year={2020},
  publisher={APS},
}

@Article{Chickarmane2009,
  author    = {Chickarmane, Vijay and Enver, Tariq and Peterson, Carsten},
  journal   = {PLoS Computational Biology},
  title     = {Computational Modeling of the Hematopoietic Erythroid-Myeloid Switch Reveals Insights into Cooperativity, Priming, and Irreversibility},
  year      = {2009},
  issn      = {1553-7358},
  month     = jan,
  number    = {1},
  pages     = {e1000268},
  volume    = {5},
  comment   = {quite a lot of other articles reference this one
PU-GATA switch
I think it's the basis of the Strasser article although I don't actually know
it's not exactly a toggle switch but has a lot of related mechanisms},
  doi       = {10.1371/journal.pcbi.1000268},
  editor    = {Antia, Rustom},
  publisher = {Public Library of Science (PLoS)},
}

@Article{Schwanhaeusser2011,
  author    = {Schwanhäusser, Björn and Busse, Dorothea and Li, Na and Dittmar, Gunnar and Schuchhardt, Johannes and Wolf, Jana and Chen, Wei and Selbach, Matthias},
  journal   = {Nature},
  title     = {Global quantification of mammalian gene expression control},
  year      = {2011},
  issn      = {1476-4687},
  month     = may,
  number    = {7347},
  pages     = {337--342},
  volume    = {473},
  doi       = {10.1038/nature10098},
  publisher = {Springer Science and Business Media LLC},
}

@Article{Gillespie1976,
  author   = {Daniel T Gillespie},
  journal  = {Journal of Computational Physics},
  title    = {A general method for numerically simulating the stochastic time evolution of coupled chemical reactions},
  year     = {1976},
  issn     = {0021-9991},
  number   = {4},
  pages    = {403-434},
  volume   = {22},
  doi      = {https://doi.org/10.1016/0021-9991(76)90041-3},
  url      = {https://www.sciencedirect.com/science/article/pii/0021999176900413},
}

@Article{Lipshtat2006,
  author    = {Lipshtat, Azi and Loinger, Adiel and Balaban, Nathalie Q. and Biham, Ofer},
  journal   = {Physical Review Letters},
  title     = {Genetic Toggle Switch without Cooperative Binding},
  year      = {2006},
  issn      = {1079-7114},
  month     = may,
  number    = {18},
  pages     = {188101},
  volume    = {96},
  doi       = {10.1103/physrevlett.96.188101},
  publisher = {American Physical Society (APS)},
}

@Book{Strogatz2018,
  author    = {Strogatz, Steven H.},
  publisher = {CRC Press},
  title     = {Nonlinear Dynamics and Chaos},
  year      = {2018},
  isbn      = {9780429961113},
  month     = may,
  doi       = {10.1201/9780429492563},
}

@Article{Gardner2000,
  author    = {Timothy S. Gardner, Charles R. Cantor, James J. Collins},
  journal   = {Nature},
  title     = {Construction of a genetic toggle switch in Escherichia coli},
  year      = {2000},
  issn      = {1476-4687},
  month     = jan,
  number    = {6767},
  pages     = {339-342},
  volume    = {403},
  comment   = {original formulation of the model, experimental and theoretical model of the network},
  doi       = {https://doi.org/10.1038/35002131},
  publisher = {Springer Science and Business Media LLC},
}

@Article{Cinquin2005,
  author    = {Cinquin, Olivier and Demongeot, Jacques},
  journal   = {Journal of Theoretical Biology},
  title     = {High-dimensional switches and the modelling of cellular differentiation},
  year      = {2005},
  issn      = {0022-5193},
  month     = apr,
  number    = {3},
  pages     = {391--411},
  volume    = {233},
  doi       = {10.1016/j.jtbi.2004.10.027},
  publisher = {Elsevier BV},
}

@Article{Duff2011,
  author    = {Duff, Campbell and Smith-Miles, Kate and Lopes, Leo and Tian, Tianhai},
  journal   = {Journal of Mathematical Biology},
  title     = {Mathematical modelling of stem cell differentiation: the PU.1–GATA-1 interaction},
  year      = {2011},
  issn      = {1432-1416},
  month     = apr,
  number    = {3},
  pages     = {449--468},
  volume    = {64},
  doi       = {10.1007/s00285-011-0419-3},
  publisher = {Springer Science and Business Media LLC},
}

@Article{Strasser2012,
  author    = {Strasser, Michael and Theis, Fabian J. and Marr, Carsten},
  journal   = {Biophysical Journal},
  title     = {Stability and Multiattractor Dynamics of a Toggle Switch Based on a Two-Stage Model of Stochastic Gene Expression},
  year      = {2012},
  issn      = {0006-3495},
  month     = jan,
  number    = {1},
  pages     = {19--29},
  volume    = {102},
  comment   = {one of the few examples where the toggle switch is written in terms of its elementary reactions. This is specifically for the case of no cooperativity (monomeric), i.e Hill coefficient = 1 for both genes.},
  doi       = {10.1016/j.bpj.2011.11.4000},
  publisher = {Elsevier BV},
}

@Article{Loinger2007,
  author    = {Loinger, Adiel and Lipshtat, Azi and Balaban, Nathalie Q. and Biham, Ofer},
  journal   = {Physical Review E},
  title     = {Stochastic simulations of genetic switch systems},
  year      = {2007},
  issn      = {1550-2376},
  month     = feb,
  number    = {2},
  pages     = {021904},
  volume    = {75},
  comment   = {Slightly more fundamental approach to the formulation of the system, helped me to understand the QSSA as it derives it more explicitly which in turn somewhat explains the form of the Hill Equation, especially for a repressor.
Maybe the first/most thorough stochastic simulation of the system? (not checked this at all)},
  doi       = {10.1103/physreve.75.021904},
  publisher = {American Physical Society (APS)},
}

@Article{Oppenheim2005,
  author    = {Oppenheim, Amos B. and Kobiler, Oren and Stavans, Joel and Court, Donald L. and Adhya, Sankar},
  journal   = {Annual Review of Genetics},
  title     = {Switches in Bacteriophage Lambda Development},
  year      = {2005},
  issn      = {1545-2948},
  month     = dec,
  number    = {1},
  pages     = {409--429},
  volume    = {39},
  doi       = {10.1146/annurev.genet.39.073003.113656},
  publisher = {Annual Reviews},
}

@Article{Li2020,
  author           = {Li, Jie and Zhang, Weinian},
  journal          = {Discrete \& Continuous Dynamical Systems - B},
  title            = {Transition between monostability and bistability of a genetic toggle switch in Escherichia coli},
  year             = {2020},
  issn             = {1553-524X},
  number           = {5},
  pages            = {1871--1894},
  volume           = {25},
  comment          = {Explains where the labels on the Gardner bistability diagram come from (i.e. quite a lot of approximations)

simplify the model (s.t. beta=1, gamma=n)
rearrange for 1 equation only involving u
consider 0<u<a
consider n=1 (i.e. a quadratic) 
-> one stable node
the momment n>1, too complicated to solve
consider the first and second derivatives of the polynomial for n
find the value(s) of u that give the second derivative = 0
find the sign of the third derivative at this point -> shows it minimizes the first derivative
use this value of u and set the first derivative equal to 0 to find conditions on a (alpha in the original model)},
  comment-s2674429 = {Explains where the labels on the Gardner bistability diagram come from (i.e. quite a lot of approximations)},
  doi              = {10.3934/dcdsb.2020007},
  publisher        = {American Institute of Mathematical Sciences (AIMS)},
}

@Article{Lu2004,
  author    = {Lu, T. and Volfson, D. and Tsimring, L. and Hasty, J.},
  journal   = {Systems Biology},
  title     = {Cellular growth and division in the Gillespie algorithm},
  year      = {2004},
  issn      = {1741-248X},
  month     = jun,
  number    = {1},
  pages     = {121--128},
  volume    = {1},
  doi       = {10.1049/sb:20045016},
  publisher = {Institution of Engineering and Technology (IET)},
}

@Article{Bierbaum2015,
  author    = {Bierbaum, Veronika and Klumpp, Stefan},
  journal   = {Physical Biology},
  title     = {Impact of the cell division cycle on gene circuits},
  year      = {2015},
  issn      = {1478-3975},
  month     = sep,
  number    = {6},
  pages     = {066003},
  volume    = {12},
  comment   = {Fundamental paper on modelling cell division},
  doi       = {10.1088/1478-3975/12/6/066003},
  publisher = {IOP Publishing},
}

@Article{Perkins2009,
  author    = {Perkins, Theodore J and Swain, Peter S},
  journal   = {Molecular Systems Biology},
  title     = {Strategies for cellular decision‐making},
  year      = {2009},
  issn      = {1744-4292},
  month     = Nov,
  number    = {1},
  volume    = {5},
  doi       = {10.1038/msb.2009.83},
  publisher = {Springer Science and Business Media LLC},
}

@Article{Balazsi2011,
  author    = {Balázsi, Gábor and van Oudenaarden, Alexander and Collins, James J.},
  journal   = {Cell},
  title     = {Cellular Decision Making and Biological Noise: From Microbes to Mammals},
  year      = {2011},
  issn      = {0092-8674},
  month     = Mar,
  number    = {6},
  pages     = {910--925},
  volume    = {144},
  doi       = {10.1016/j.cell.2011.01.030},
  publisher = {Elsevier BV},
}

@Article{Maizels2026,
  author    = {Maizels, Rory J. and Briscoe, James},
  journal   = {Nature Reviews Genetics},
  title     = {Gene regulatory networks: from correlative models to causal explanations},
  year      = {2026},
  issn      = {1471-0064},
  month     = Mar,
  number    = {6},
  pages     = {485--498},
  volume    = {27},
  doi       = {10.1038/s41576-026-00939-1},
  publisher = {Springer Science and Business Media LLC},
}

@Article{Biancalani2014,
  author    = {Biancalani, Tommaso and Dyson, Louise and McKane, Alan J.},
  journal   = {Physical Review Letters},
  title     = {Noise-Induced Bistable States and Their Mean Switching Time in Foraging Colonies},
  year      = {2014},
  issn      = {1079-7114},
  month     = Jan,
  number    = {3},
  pages     = {038101},
  volume    = {112},
  doi       = {10.1103/physrevlett.112.038101},
  publisher = {American Physical Society (APS)},
}

@Article{Santillan2008,
  author    = {Santillán, M.},
  journal   = {Mathematical Modelling of Natural Phenomena},
  title     = {On the Use of the Hill Functions in Mathematical Models of Gene Regulatory Networks},
  year      = {2008},
  issn      = {0973-5348},
  number    = {2},
  pages     = {85--97},
  volume    = {3},
  doi       = {10.1051/mmnp:2008056},
  publisher = {EDP Sciences},
}

@Article{PerezCarrasco2016,
  author    = {Perez-Carrasco, Ruben and Guerrero, Pilar and Briscoe, James and Page, Karen M.},
  journal   = {PLOS Computational Biology},
  title     = {Intrinsic Noise Profoundly Alters the Dynamics and Steady State of Morphogen-Controlled Bistable Genetic Switches},
  year      = {2016},
  issn      = {1553-7358},
  month     = Oct,
  number    = {10},
  pages     = {e1005154},
  volume    = {12},
  doi       = {10.1371/journal.pcbi.1005154},
  editor    = {Tusscher, Ten},
  publisher = {Public Library of Science (PLoS)},
}

@Article{Lee2018,
  author    = {Lee, Sangmi and Lewis, Dale E.A. and Adhya, Sankar},
  journal   = {Journal of Molecular Biology},
  title     = {The Developmental Switch in Bacteriophage $\lambda$: A Critical Role of the Cro Protein},
  year      = {2018},
  issn      = {0022-2836},
  month     = Jan,
  number    = {1},
  pages     = {58--68},
  volume    = {430},
  doi       = {10.1016/j.jmb.2017.11.005},
  publisher = {Elsevier BV},
}

@Article{Atsumi2006,
  author    = {Atsumi, Shota and Little, John W.},
  journal   = {Proceedings of the National Academy of Sciences},
  title     = {A synthetic phage $\lambda$ regulatory circuit},
  year      = {2006},
  issn      = {1091-6490},
  month     = Dec,
  number    = {50},
  pages     = {19045--19050},
  volume    = {103},
  doi       = {10.1073/pnas.0603052103},
  publisher = {Proceedings of the National Academy of Sciences},
}

@Article{Tian2004,
  author    = {Tian, Tianhai and Burrage, Kevin},
  journal   = {Journal of Theoretical Biology},
  title     = {Bistability and switching in the lysis/lysogeny genetic regulatory network of bacteriophage $\lambda$},
  year      = {2004},
  issn      = {0022-5193},
  month     = Mar,
  number    = {2},
  pages     = {229--237},
  volume    = {227},
  doi       = {10.1016/j.jtbi.2003.11.003},
  publisher = {Elsevier BV},
}

@Article{Biswas2022,
  author    = {Biswas, Kuheli and Jolly, Mohit Kumar and Ghosh, Anandamohan},
  journal   = {Journal of Biosciences},
  title     = {Mean residence times of TF-TF and TF-miRNA toggle switches},
  year      = {2022},
  issn      = {0973-7138},
  month     = Apr,
  number    = {2},
  volume    = {47},
  doi       = {10.1007/s12038-022-00261-y},
  publisher = {Springer Science and Business Media LLC},
}

@Article{Stefan2013,
  author    = {Stefan, Melanie I. and Le Novère, Nicolas},
  journal   = {PLoS Computational Biology},
  title     = {Cooperative Binding},
  year      = {2013},
  issn      = {1553-7358},
  month     = Jun,
  number    = {6},
  pages     = {e1003106},
  volume    = {9},
  doi       = {10.1371/journal.pcbi.1003106},
  editor    = {Wodak, Shoshana},
  publisher = {Public Library of Science (PLoS)},
}

@Article{PerezOrtin2019,
  author    = {Pérez-Ortín, José E. and Tordera, Vicente and Chávez, Sebastián},
  journal   = {RNA Biology},
  title     = {Homeostasis in the Central Dogma of molecular biology: the importance of mRNA instability},
  year      = {2019},
  issn      = {1555-8584},
  month     = Sep,
  number    = {12},
  pages     = {1659--1666},
  volume    = {16},
  doi       = {10.1080/15476286.2019.1655352},
  publisher = {Informa UK Limited},
}

@Article{Voichek2016,
  author    = {Voichek, Yoav and Bar-Ziv, Raz and Barkai, Naama},
  journal   = {Science},
  title     = {Expression homeostasis during DNA replication},
  year      = {2016},
  issn      = {1095-9203},
  month     = Mar,
  number    = {6277},
  pages     = {1087--1090},
  volume    = {351},
  doi       = {10.1126/science.aad1162},
  publisher = {American Association for the Advancement of Science (AAAS)},
}

@Book{Mezoe2022,
  author    = {Mező, István},
  publisher = {Chapman and Hall/CRC},
  title     = {The Lambert W Function: Its Generalizations and Applications},
  year      = {2022},
  isbn      = {9781003168102},
  month     = Mar,
  doi       = {10.1201/9781003168102},
}


\appendix

\setcounter{figure}{0}
\renewcommand{\thefigure}{A\arabic{figure}}

\section{Reaction Scheme for the Toggle Switch}

The toggle switch motif can be described via the following reaction scheme:
\begin{align}
    &G_A \xrightarrow{\alpha_A} G_A + P_A, \label{eqn:A_synthesis} \\
    &G_B \xrightarrow{\alpha_B} G_B + P_B, \label{eqn:B_synthesis} \\
    &G_A + h_1P_B  \xrightleftharpoons[k_{A-}]{k_{A+}} G_A^*, \label{eqn:A_binding} \\
    &G_B + h_2P_A  \xrightleftharpoons[k_{B-}]{k_{B+}} G_B^*, \label{eqn:B_binding} \\
    &P_A \xrightarrow{\beta_A + \delta} \emptyset, \label{eqn:A_degradation} \\
    &P_B \xrightarrow{\beta_B + \delta} \emptyset, \label{eqn:B_degradation}
\end{align}
where the reactions in \eqref{eqn:A_synthesis} and \eqref{eqn:B_synthesis} describe the synthesis of proteins A and B from their respective genes with rates $\alpha_A$ and $\alpha_B$; the reactions in \eqref{eqn:A_binding} and \eqref{eqn:B_binding} represent the reversible switching of each of the promoters from their active to their inactive states via the cooperative binding and unbinding of the opposing protein, with rates $k_{X+}$ and $k_{X-}$ ($X \in \{A,B\}$); and \eqref{eqn:A_degradation} and \eqref{eqn:B_degradation} describe the active degradation and dilution of each protein, with rates $\beta_A + \delta$ and $\beta_B + \delta$, respectively.

It is commonly assumed that the binding reactions in \eqref{eqn:A_binding} and \eqref{eqn:B_binding} occur on a much faster timescale than the remaining reactions. A quasi-steady-state approximation (QSSA) can then be applied to obtain a reduced model in which Hill functions are introduced to account for cooperative protein-promoter binding \cite{Stefan2013,Santillan2008}, which results in the system of ODEs in Eqs.~\eqref{eqn:NA_classical} and \eqref{eqn:NB_classical}. Here, the threshold concentrations $\lambda_A$ and $\lambda_B$ derive from the dissociation constants $K_A = k_{A-}/k_{A+}$ and $K_B = k_{B-}/k_{B+}$, respectively, such that $K_A = \lambda_B^{h_1}$ and $K_B = \lambda_A^{h_2}$, where the total copy number for each gene is assumed to be $1$.

\section{Derivation of the Boolean Toggle Switch (BTS)}

To achieve the binarisation of the activation function in the Boolean toggle switch (BTS), let us consider Eqs. \eqref{eqn:NA_classical} and \eqref{eqn:NB_classical} from the Main Text in the limit of infinite Hill coefficients.

Let $\phi_X = \frac{N_X}{\lambda_XV}$ for $X \in \{A,B\}$, and take $h_1, h_2 \to \infty$:

$$
\lim_{h_1\to\infty} \frac{\alpha_A}{1+\phi_B^{h_1}}=\begin{cases}
            0 & \text{if $\phi_B > 1 \Leftrightarrow N_B > \lambda_B V$}, \\
            \alpha_A & \text{if $\phi_B < 1 \Leftrightarrow N_B < \lambda_B V$.}
		 \end{cases}
$$
Similarly,
$$
\lim_{h_2\to\infty} \frac{\alpha_B}{1+\phi_A^{h_2}}=\begin{cases}
            0 & \text{if $\phi_A > 1 \Leftrightarrow N_A > \lambda_A V$}, \\
            \alpha_B & \text{if $\phi_A < 1 \Leftrightarrow N_A < \lambda_A V$.}
		 \end{cases}
$$
Hence, we see that these Hill functions can be replaced by Heaviside step functions in that limit, giving us fixed concentration thresholds for the repression of one gene by the other in the network.


\section{Separatrix derivations for non-zero protein degradation}

\begin{figure}
    \centering
    \includegraphics[width=0.5\linewidth]{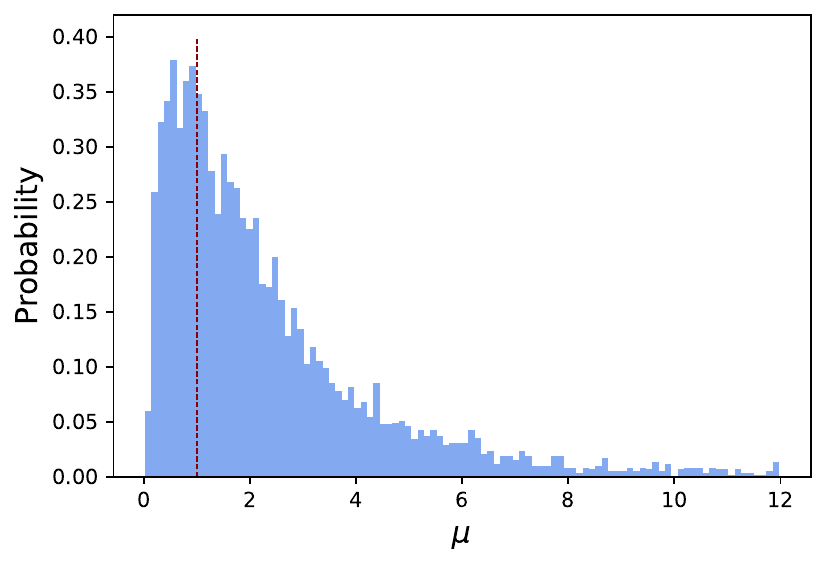}
    \caption{Distribution of $\mu$, the ratio of protein \textcolor{black}{half-life -- defined as $\ln 2$ divided by the protein degradation rate --} to median cell cycle duration, derived from data in Ref.~\cite{Schwanhaeusser2011}. The red line marks $\mu = 1$; to the right of that line, protein degradation is negligible. Data is taken from 5028 proteins from mouse fibroblast cells, with a median cell cycle duration of 27.5 hours.}
    \label{fig:mu_dist}
\end{figure}

As discussed in the Main Text, from Ref.~\cite{Schwanhaeusser2011} we find that approximately 71$\%$ of proteins in mammalian cells have a half-life that is longer than the median cell cycle length. If $\mu$ is the ratio of protein lifetime to cell cycle length, then for proteins with a $\mu$-value above $1$, the effect of protein degradation is negligible during the cell cycle; thus, we can set $\beta_A=0=\beta_B$. Fig.~\ref{fig:mu_dist} illustrates the distribution of $\mu$-values resulting from Ref.~\cite{Schwanhaeusser2011}. 

For proteins with a $\mu$-value below $1$, i.e. with half-lives that are shorter than the cell cycle length, active protein degradation is no longer negligible and must be included in our modelling. We hence repeat the separatrix derivations from the Main Text here for non-zero degradation rates $\beta_A, \beta_B > 0$.

We again first consider the standard BTS in the two cases where (i) $A_0 < \lambda_A$, $B_0<\lambda_B$ and (ii) $A_0 \geq\lambda_A$, $B_0 \geq \lambda_B$. Recall that the remaining cases ($A_0 < \lambda_A$, $B_0\geq\lambda_B$; $A_0 \geq \lambda_A$, $B_0 < \lambda_B$) do not contribute to the separatrix.

In case (i), the time at which each protein will reach its repression threshold is given by
\begin{equation*}
    t_{\lambda_X} = \frac{1}{\beta_X+\delta}\ln\frac{\alpha_X - (\beta_X+\delta)X_0}{\alpha_X - (\beta_X +\delta)\lambda_X}.
\end{equation*}
Hence, the separatrix is defined by $t_{\lambda_A} = t_{\lambda_B}$ (segment $\hat{S}_1$; Table~\ref{tab:non_zero_degradation_separatrix_expressions}, row 1). 

The separatrix for case (ii) is given by segment $\hat{S}_2$ (Table~\ref{tab:non_zero_degradation_separatrix_expressions}, row 2).

We then consider the BTS with division, again in the two separate cases where (i) $A_0 < \lambda_AV_0$, $B_0<\lambda_BV_0$ and (ii) $A_0 \geq\lambda_AV_0$, $B_0 \geq \lambda_BV_0$. 

In order to determine the separatrix in case (i), we need to find $t$ such that
\begin{equation*}
    \frac{\alpha_X}{\beta_X} + \left(X_0 - \frac{\alpha_X}{\beta_X}\right)e^{-\beta_Xt} = \lambda_XV_0e^{\delta t}
\end{equation*}
which, however, does not seem to admit a closed-form solution.
Instead, we again make use of the linear approximation $\tilde{\psi}_X(t) = \lambda_XV_0(1 + t/T)$ for the protein threshold $\psi_X(t)$, which leads to
\begin{equation}
    t_{\psi_X} = T\left(\frac{\alpha_X}{\beta_X\lambda_XV_0}-1\right) \\ +\frac{1}{\beta_X}W_{k}\left(\frac{\beta_X T}{\lambda_X V_0}\left(X_0 - \frac{\alpha_X}{\beta_X}\right)\exp\left(\beta_X T - \frac{\alpha_X T}{\lambda_X V_0}\right)\right).
\end{equation}
Here, $W_{k}$ is an appropriately chosen branch of the Lambert W function, with $k\in\mathbb{Z}$ \cite{Mezoe2022}. An approximation to the separatrix is then given by $t_{\psi_A} = t_{\psi_B}$ (segment $\hat{L}_1$; Table~\ref{tab:non_zero_degradation_separatrix_expressions}, rows 5 and 6).

For case (ii), the separatrix is defined by segment $\hat{D}_2$ (Table \ref{tab:non_zero_degradation_separatrix_expressions}, row 4), derived by the same threshold crossing argument.

\textcolor{black}{We are not aware of a limit in which the separatrix for the BTS with division converges to that for the standard BTS. Even in the limit of fast active protein degradation, $\beta_X \rightarrow \infty$ ($X\in\{A,B\}$), the BTS with division predicts time-varying concentrations within each cell cycle, whereas the standard BTS always predicts constant concentrations in the limit of long times. This discrepancy can be traced to the fact that the Heaviside functions in Eqs.~\eqref{eqn:NA_BTS} and \eqref{eqn:NB_BTS} involve a dependence on volume -- and, hence, on time -- whereas those in Eqs.~\eqref{eqn:NA_standard_BTS} and \eqref{eqn:NB_standard_BTS} do not. }

\renewcommand{\arraystretch}{2}
\begin{table}
    \centering
        \begin{tabular}{ccc}
        \hline \hline
        Model & Segment label & Separatrix expression \\
        \hline
        Non-zero degradation & & \\
        Standard BTS & $\hat{S}_1$ & $\frac{1}{\beta_A+\delta}\ln\frac{\alpha_A - (\beta_A+\delta)A_0}{\alpha_A - (\beta_A +\delta)\lambda_A} = \frac{1}{\beta_B+\delta}\ln\frac{\alpha_B - (\beta_B+\delta)B_0}{\alpha_B - (\beta_B +\delta)\lambda_B}
        $ \\
      & $\hat{S}_2$ & $\frac{1}{\beta_A+\delta}\ln\frac{A_0}{\lambda_A} = \frac{1}{\beta_B+\delta}\ln\frac{B_0}{\lambda_B}$ \\
      BTS with division & $\hat{D}_1$ & Analytically intractable \\
        & $\hat{D}_2$ &  $\frac{1}{\beta_A+\delta}\ln\left(\frac{A_0}{\lambda_A V_0}\right) = \frac{1}{\beta_B+\delta}\ln\left(\frac{B_0}{\lambda_B V_0}\right)$ \\
    Linear approximation & $\hat{L}_1$ & $T\big(\frac{\alpha_A}{\beta_A\lambda_AV_0}-1\big) +\frac{1}{\beta_A}W_{k}\left(\frac{\beta_A T}{\lambda_A V_0}\big(A_0 - \frac{\alpha_A}{\beta_A}\big)\exp\big(\beta_A T - \frac{\alpha_A T}{\lambda_A V_0}\big)\right)$ \\
    & & $= T\big(\frac{\alpha_B}{\beta_B\lambda_BV_0}-1\big) +\frac{1}{\beta_B}W_{k}\left(\frac{\beta_B T}{\lambda_B V_0}\big(B_0 - \frac{\alpha_B}{\beta_B}\big)\exp\big(\beta_B T - \frac{\alpha_B T}{\lambda_B V_0}\big)\right)$ \\
    \hline \hline
    \end{tabular}
    
    \caption{Expressions for the separatrices for both the standard BTS and the BTS with division, in the case of non-zero active degradation ($\beta_A,\beta_B > 0$).}
    \label{tab:non_zero_degradation_separatrix_expressions}
\end{table}

\section{Bistability conditions for the standard BTS}

The following analysis applies to the standard BTS with general degradation rates $\beta_A,\beta_B\geq 0$. In the Main Text, the zero-degradation case where $\beta_A=0=\beta_B$ is considered, for which the bistability condition becomes $\alpha_X/\delta > \lambda_X$.

In general, if $\frac{\alpha_A}{\beta_A + \delta} <\lambda_A$ or $\frac{\alpha_B}{\beta_B + \delta} <\lambda_B$, the standard BTS is a monostable system. To understand why, recall that each protein is essentially either abiding by a birth-death process ($H(x) = 1$) or by a pure death process ($H(x) = 0$), depending on the count of the opposing protein. For a birth-death process
\begin{equation}
    \frac{\text{d}N_X}{\text{dt}} = \alpha_X - (\beta_X + \delta)N_X,
\end{equation}
the steady state reached is $N_X^\ast = \alpha_X/(\beta_X + \delta)$. A pure death process assumes its steady state at $N_X^\ast=0$. 

Now, assume $\frac{\alpha_A}{\beta_A + \delta} <\lambda_A$ and $\frac{\alpha_B}{\beta_B + \delta} >\lambda_B$. If $A_0 < \lambda_A$ and $B_0 < \lambda_B$, then both proteins are abiding by the full birth-death dynamics. However, $N_A$ will reach steady state at $\alpha_A/(\beta_A+\delta)$ which, by our assumption, is below the threshold $\lambda_A$, meaning that synthesis of protein B will always continue, allowing B to reach its threshold, switching A OFF and allowing B to ``win". Similarly, if $A_0 > \lambda_A$ and $B_0 < \lambda_B$, protein A will decrease even though its synthesis is ON until it reaches its steady-state level which is below the threshold, once again switching synthesis of protein B ON, resulting in a steady state where protein B ``wins". If $A_0 < \lambda_A$ and $B_0 > \lambda_B$, B will win trivially. If both $A_0 > \lambda_A$ and $B_0 > \lambda_B$, both proteins will be abiding by a pure death process and therefore decrease to 0. However, as soon as either reaches its threshold, we will find ourselves in a scenario that is equivalent to one of the two cases previously discussed. Hence, there will only ever be one steady state where protein B ``wins" if $\frac{\alpha_A}{\beta_A + \delta} <\lambda_A$ and $\frac{\alpha_B}{\beta_B + \delta} >\lambda_B$.

Likewise, there will only ever be a steady state where protein A wins if $\frac{\alpha_A}{\beta_A+\delta} >\lambda_A$ and $\frac{\alpha_B}{\beta_B+\delta} <\lambda_B$.

Following the same logic, it can be shown that if both  $\frac{\alpha_A}{\beta + \delta} <\lambda_A$ and $\frac{\alpha_B}{\beta + \delta} <\lambda_B$, regardless of initial conditions, the system will always reach a steady state in which both proteins have reached their birth-death steady state levels $N_X^\ast = \alpha_X/(\beta_X + \delta)$ ($X\in\{A,B\}$). 

Hence, if either $\frac{\alpha_A}{\beta_A + \delta} <\lambda_A$ or $\frac{\alpha_B}{\beta_B+\delta} <\lambda_B$, the system is monostable.

\begin{figure}
    \centering
    \begin{overpic}[width=0.75\linewidth]{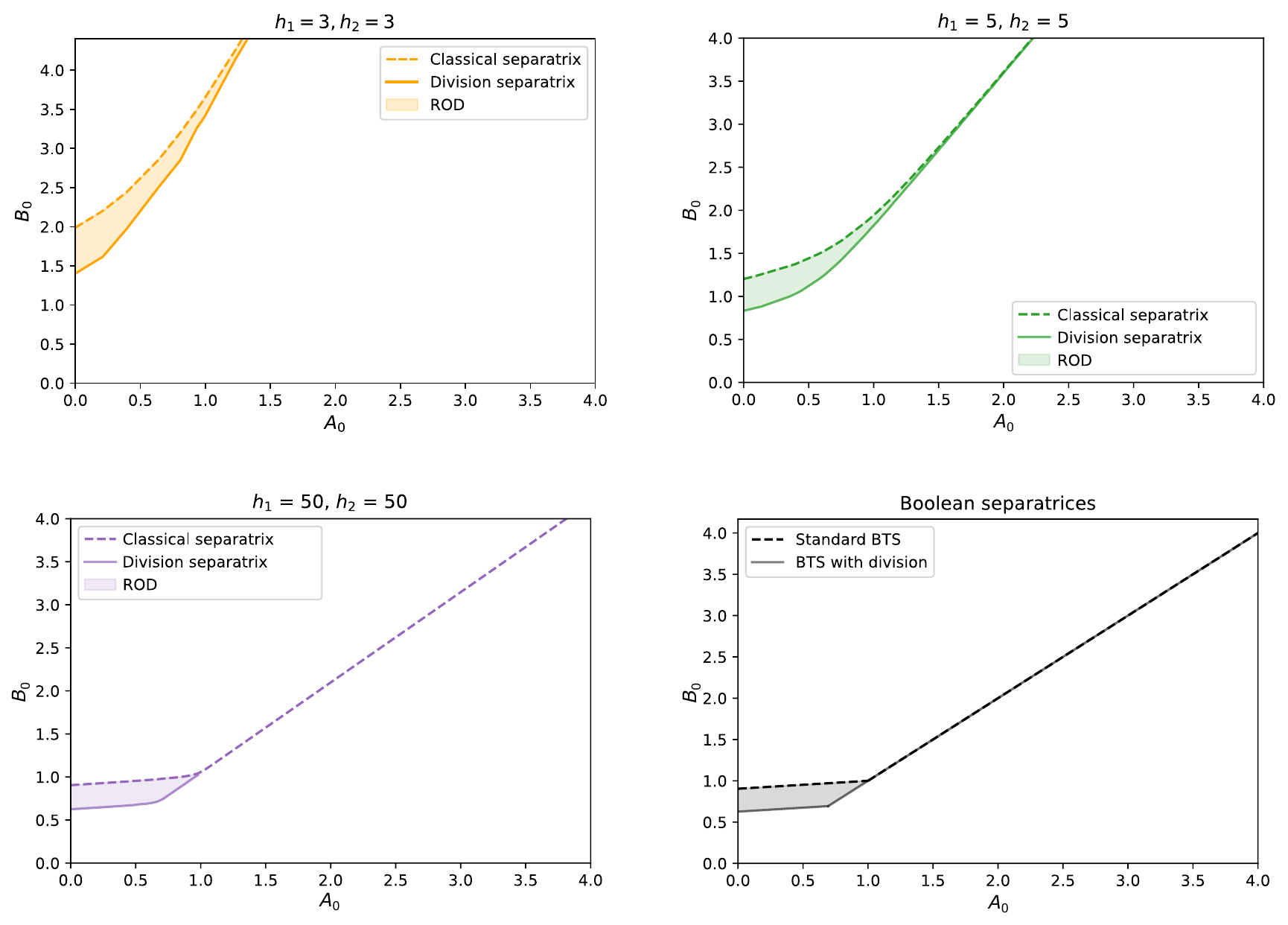}
    \put(0,705){(a)}
    \put(510,705){(b)}
    \put(0,340){(c)}
    \put(510,340){(d)}
    \end{overpic}
    \caption{\textcolor{black}{Plots of the ROD for the classical toggle switch model in Eqs.~\eqref{eqn:NA_classical} and \eqref{eqn:NB_classical} with $V = 1$ and its cell-division extension, obtained numerically for Hill coefficients (a) $h_1=3=h_2$, (b) $h_1=5=h_2$, and (c) $h_1=50=h_2$. Note that the version of the classical toggle switch with cell division involves the numerical solution of Eqs.~\eqref{eqn:NA_classical} and \eqref{eqn:NB_classical} with $\delta = 0$, $V = V_0 e^{\frac{\ln 2}{T}t}$, and $t \in [0,T]$, where $t$ is the time within the cell cycle; when $t = T$, the protein count is halved and $t$ is reset to $0$. (d) The ROD for the BTS. Note that the ROD for the classical model converges to that for the BTS with increasing Hill coefficients. Throughout, $\alpha_A = 150$, $\alpha_B = 15$, $\beta_A = 0 = \beta_B$, $\lambda_A = 1 = \lambda_B$, and $T = 27.5$.}}
    \label{fig:finite_hill_BTS}
\end{figure}

\section{Separatrix derivation for the BTS with division}

In deriving expressions for the separatrix for the BTS with division, we follow the same procedure as for the standard BTS: we consider two separate cases; in both of those, we determine the point in time for which both protein counts reach their repression thresholds simultaneously. We assume $\alpha_X T > 2V_0\lambda_X$ to ensure that both proteins can reach their thresholds within a single cell cycle. Recall that for the BTS with division, cell growth is modelled exponentially, with the protein count repression threshold, $\psi_X(t)$, proportional to the cell volume:
\begin{equation*}
    \psi_X(t) = \lambda_XV_0e^{\delta t}.
\end{equation*}

For case (i) ($A_0 < \lambda_A V_0$, $B_0<\lambda_B V_0$), we can find $t_{\psi_X}$ by solving, then rearranging, the following expression,
\begin{equation}
    X_0 +\alpha_Xt_{\psi_X} = V_0\lambda_Xe^{\delta t_{\psi_X}},
\end{equation}
which leads to
\begin{equation}
    t_{\psi_X} = - \frac{X_0}{\alpha_X} - \frac{W_0\Big(-\frac{\delta \lambda_XV_0 \exp{\big(-\frac{\delta}{\alpha_X}X_0\big)}}{\alpha_X}\Big)}{\delta}.
\end{equation}
Here, $W_0$ is the principal branch of the Lambert W function; \textcolor{black}{see subsection E.1 below for validation.}

Hence, the separatrix is given by

\begin{equation}
    \frac{A_0}{\alpha_A} + \frac{W_0\Big(-\frac{\delta \lambda_A V_0 \exp{\big(-\frac{\delta}{\alpha_A}A_0\big)}}{\alpha_A}\Big)}{\delta} \\ = \frac{B_0}{\alpha_B} + \frac{W_0\Big(-\frac{\delta \lambda_B V_0 \exp{\big(-\frac{\delta}{\alpha_B}B_0\big)}}{\alpha_B}\Big)}{\delta}. \label{eqn:division_sep_1}
\end{equation}
In the Main Text, the simplified notation 
\begin{equation}
    \Omega_X = W_0\left(-\frac{\delta \lambda_X V_0 \exp{\big(-\frac{\delta}{\alpha_X}X_0\big)}}{\alpha_X}\right)
    \label{eqn:omega}
\end{equation}
is used for clarity and brevity.

 Given the algebraic complexity of Eq.~\eqref{eqn:division_sep_1}, we also formulate a simplified expression for the resulting separatrix, via a linear approximation for the protein count threshold; \textcolor{black}{see subsection E.2 below}. That simplification allows us to investigate the geometry of the separatrices and approximate the area of the ROD, as the Lambert W function in Eq.~\eqref{eqn:division_sep_1} precludes a straightforward geometrical analysis.

For case (ii) ($A_0 \geq \lambda_AV_0$, $B_0\geq\lambda_BV_0$), we again repeat the same procedure to find $t_{\psi_X}$: $X_0 = \lambda_XV_0 e^{\delta t_{\psi_X}}$ implies
\begin{align*}
    t_{\psi_X} = \frac{1}{\delta}\ln\frac{X_0}{\lambda_XV_0}.
\end{align*}
Note that the argument of the logarithm is always positive, given that $X_0, \lambda_X, V_0>0$.

Hence, the separatrix is given by
\begin{equation}
    \frac{A_0}{\lambda_A} = \frac{B_0}{\lambda_B}. \label{eqn:division_sep_2}
\end{equation}

For cases (iii) and (iv) ($A_0< \lambda_AV_0$, $B_0 \geq \lambda_BV_0$; $A_0 \geq \lambda_AV_0$, $B_0 < \lambda_BV_0$), as in the standard BTS, there is no case for competition since whichever protein has its count above its threshold will win. 

\subsection{\textcolor{black}{The Lambert W function for zero protein degradation}}

The derivation of segment $D_1$ for the case of zero degradation in the Main Text makes use of the Lambert W function, $W(z)$, which is defined as the solution of the equation $We^W=z$; it has one real solution on the $z$-interval $[0,\infty)$ and two real solutions on the $z$-interval $(-e^{-1},0)$ \cite{Mezoe2022}.

Eq.~\eqref{eqn:division_sep_1} implies that the Lambert W function appears with the following argument,
\begin{equation*}
    z = -\frac{\delta \lambda_X V_0\exp(-\frac{\delta}{\alpha_X}X_0)}{\alpha_X}.
\end{equation*}
Since $X_0,\alpha_X,\delta \geq 0$, it follows that $z<0$ and $0<\exp(-\frac{\delta}{\alpha_X}X_0) \leq 1$. From our assumption that $\alpha_X T > 2V_0 \lambda_X$, we find
\begin{equation}
    |z| = \frac{\delta \lambda_X V_0}{\alpha_X}\exp(-\tfrac{\delta}{\alpha_X}X_0) \leq \frac{\delta \lambda_X V_0}{\alpha_X} \\ = \frac{\ln2 \lambda_X V_0}{\alpha_X T} < \frac{\ln2 \lambda_X V_0}{2 V_0 \lambda_X} = \frac{\ln 2}{2} < \frac{1}{e}.
    \label{eqn:z_condition}
\end{equation}
Hence, our solution certainly lies on a real branch of the Lambert W function.

To determine which of the real branches we must select -- the principal branch $W_0$ or the secondary branch $W_{-1}$ -- we first note that $W_0(z) \geq -1$ and $W_{-1}(z)\leq-1$. 

If we consider the time at which each protein reaches its threshold, the inequality $0 < t_{\psi_X} < T$ must be satisfied in order for the threshold to be reached during the first cycle. Hence,
\begin{equation}
    0 < -\frac{X_0}{\alpha_X} - \frac{\Omega_X}{\delta} <T,
\end{equation}
where $\Omega_X = W_k(z)$ and $k\in\{-1,0\}$ is to be determined, which leads to
\begin{equation*}
    -\frac{X_0 \delta}{\alpha_X} -\ln 2 < \Omega_X < -\frac{X_0 \delta}{\alpha_X}.
\end{equation*}

Recall that this expression is only valid for $X_0<\lambda_XV_0$, given the segmentation of our separatrix derivation. Further, recall that in order for the protein count to reach its threshold in the first cycle, we require $\alpha_XT>2\lambda_XV_0$. Hence, we see that
\begin{equation*}
    0 < \frac{X_0 \delta}{\alpha_X} < \frac{\ln 2}{2},
\end{equation*}
which implies
\begin{equation}
    -\frac{3\ln2}{2} < \Omega_X<-\frac{\ln 2}{2}.
\end{equation}
Finally, recall from Eq.~\eqref{eqn:z_condition} that $|z|< \frac{\ln 2}{2}$. We see that $W_{-1}(-\frac{\ln 2}{2}) = -2\ln2 < -\frac{3 \ln2}{2} $ and, thus, that our solution requires the use of the principal branch $W_0$ of the Lambert W function.

\subsection{\textcolor{black}{Linear approximation in the BTS with division}}
Given the complexity of the expression for separatrix segment $D_1$ in the first case of the BTS with division, Eq.~\eqref{eqn:division_sep_1}, we make a linear approximation for the protein count threshold $\psi_X(t)$:
\begin{equation*}
    \tilde{\psi}_X(t) = \lambda_XV_0\left(1+\frac{t}{T}\right).
\end{equation*}
(Numerical simulation suggests that this approximation is justified for the parameter sets we consider.)

The same threshold crossing argument as in the derivations in the Main Text then yields a much simplified approximate expression for $D_1$, 
\begin{equation}
    \frac{\lambda_AV_0 - A_0}{\alpha_A - \frac{V_0\lambda_A}{T}} = \frac{\lambda_BV_0 - B_0}{\alpha_B - \frac{V_0\lambda_B}{T}}, 
    \label{eqn:div_sep_1_approx}
\end{equation}
which will be labelled $L_1$. As numerics show $L_1$ to be a good approximation to $D_1$ in all the regimes we consider, we base our discussion of the geometry of the separatrices, as well as our calculation of the area of the ROD, on $L_1$. 

\section{Geometry of the separatrices and area of the ROD}

In the Main Text, we give an expression for the approximate area of the ROD. Here, we detail the underlying calculations; we discuss the geometry of the separatrices and, in particular, the angles between the relevant segments thereof, using Eq.~\eqref{eqn:div_sep_1_approx} as an approximation to Eq.~\eqref{eqn:division_sep_1}. Recall that this approximate expression only holds for the case of zero active degradation, with $\beta_A=0=\beta_B$. 

Let us begin by rearranging the equations for separatrix segments $S_1$ and $L_1$ in standard form in order to facilitate a Euclidean analysis of their geometry. Recall that $V_0 = \ln2$ and, hence, that $V_0/T = \delta$, which gives
\begin{align}
    S_1&: B_0 = \frac{\alpha_B - \delta\lambda_B}{\alpha_A - \delta \lambda_A}A_0 + \frac{\alpha_A\lambda_B - \alpha_B\lambda_A}{\alpha_A - \delta\lambda_A}\quad\text{and} \\[0.3cm]
    L_1&: B_0 = \frac{\alpha_B - \delta\lambda_B}{\alpha_A - \delta \lambda_A}A_0 + \frac{V_0(\alpha_A\lambda_B - \alpha_B\lambda_A)}{\alpha_A - \delta\lambda_A}.
\end{align}
We see that segments $S_1$ and $L_1$ are parallel in the plane of initial conditions. Further, we recall that segments $S_2$ and $D_2$ are also parallel, as they are given by the same expression, but on different domains. Hence, the angle between $S_1$ and $S_2$ will equal that between $L_1$ and $D_2$.

We provide calculations for the parameter regime where $\alpha_A\lambda_B > \alpha_B\lambda_A$, but again highlight that a simple switching of indices ($A \leftrightarrow B$) will return expressions and limits for the opposite regime, with $\alpha_A\lambda_B < \alpha_B\lambda_A$.

\begin{figure}
    \centering
    \includegraphics[width = 0.5\linewidth]{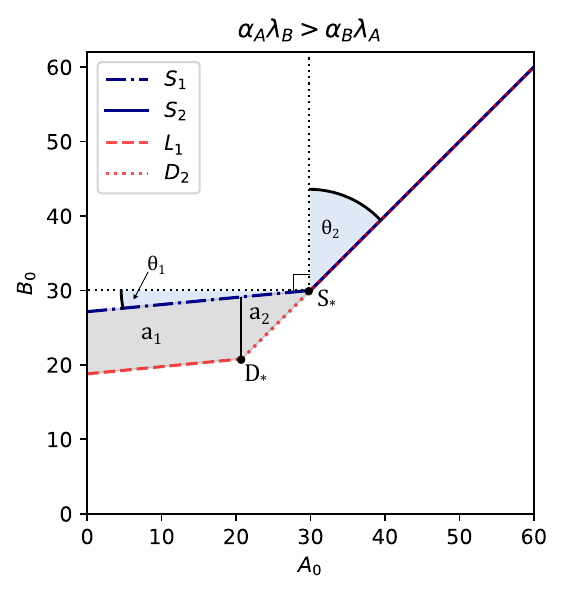}
    \caption{Schematic of the separatrices and the ROD in the plane of initial conditions, with regions required for the calculation of the angle of intersection and the area of the ROD (shaded).}
    \label{fig:area_angles}
\end{figure}

Let us denote by $S_\ast$ the point of intersection of segments $S_1$ and $S_2$ at $(\lambda_A,\lambda_B)$; correspondingly, $D_\ast$ denotes the point of intersection of segments $L_1$ and $D_2$ at $(\lambda_AV_0,\lambda_BV_0)$. Finally, let $\theta$ be the angle of intersection at both $S_\ast$ and $D_\ast$, which can be written as the sum of 3 sub-angles; see Fig.~\ref{fig:area_angles}:
\begin{enumerate}
    \item the angle $\theta_1$ between $S_1$ and the horizontal;
    \item the angle $\theta_2$ between $S_2$ and the vertical; and
    \item a $90^{\circ}$ angle between $\theta_1$ and $\theta_2$.
\end{enumerate}

From the gradients of the separatrix segments, we find
\begin{align}
    \theta_1 &= \tan^{-1}\left(\frac{\alpha_B - \delta\lambda_B}{\alpha_A - \delta\lambda_A}\right)\quad\text{and} \\
    \theta_2 &= \tan^{-1}\left(\frac{\lambda_A}{\lambda_B}\right).
\end{align}
We see that as $\alpha_A \to \infty$ ($\lambda_B \to \infty$), $\theta_1 \to 0$ ($\theta_2 \to 0$); hence, as $\alpha_A\lambda_B \to \infty$, the total angle $\theta\to 90^{\circ}$.

To calculate the area of the approximate ROD, we divide that region into two sub-regions, the parallelogram $a_1$ and the triangle $a_2$, by drawing a vertical line from $D_\ast$; see Fig.~\ref{fig:area_angles}.

Let $h$ be the vertical distance between $S_1$ and $L_1$:
\begin{equation*}
    h= \frac{\alpha_A\lambda_B - \alpha_B\lambda_A}{\alpha_A - \delta\lambda_A}(1-V_0).
\end{equation*}
The area $A_1$ of the parallelogram $a_1$ is then given by $A_1 = \lambda_AV_0 h$.

To calculate the area $A_2$ of the triangle $a_2$, let $x$ be the Euclidean distance between $D_\ast$ and $S_\ast$:
\begin{equation*}
    x = |S_\ast - D_\ast| = (1-V_0)\sqrt{\lambda_A^2 + \lambda_B^2}.
\end{equation*}
Then, $A_2$ can be calculated by $\frac{1}{2}xh\sin \theta_2$; hence, and since $\sin\theta_2=\lambda_A/\sqrt{\lambda_A^2+\lambda_B^2}$,
\begin{equation}
    A_2 = \frac{1}{2}h(1-V_0)\lambda_A.
\end{equation}

Thus, the total area $A_{\rm ROD}$ of the ROD is approximated by the sum of $A_1$ and $A_2$:
\begin{align*}
    A_{\rm ROD} &\approx \lambda_AV_0h + \frac{1}{2}h(1-V_0)\lambda_A= \frac{1}{2}\lambda_Ah(V_0 +1) \\
    &= \frac{1}{2}\lambda_A(1-V_0^2)\frac{\alpha_A\lambda_B - \alpha_B\lambda_A}{\alpha_A - \delta\lambda_A}.
\end{align*}

\section{Stochastic Simulation Algorithm (SSA) for the BTS}

Stochastic simulations were performed using the Stochastic Simulation Algorithm (SSA), modified where necessary for the inclusion of cell division \cite{Gillespie1976, Lu2004}. For both the standard BTS and the BTS with division, there are four main reactions to consider:
\begin{enumerate}
    \item[$R_1$:] the production of protein $A$;
    \item[$R_2$:] the production of protein $B$;
    \item[$R_3$:] the degradation/dilution of protein $A$; and
    \item[$R_4$:] the degradation/dilution of protein $B$.
\end{enumerate}
If $a_i$ ($i=1,\dots,4$) is the propensity associated with reaction $R_i$, for the standard BTS, we have
\begin{enumerate}
    \item $a_1 = \alpha_AH(\lambda_B - N_B)$,
    \item $a_2 = \alpha_BH(\lambda_A - N_A)$,
    \item $a_3 = (\beta_A + \delta)N_A$, and
    \item $a_4 = (\beta_B + \delta)N_B$.
\end{enumerate}
The above Heaviside functions are evaluated at the beginning of each iteration of the algorithm, with the propensities $a_1$ and $a_2$ set accordingly.

The time $\tau$ until the next reaction is then calculated from
\begin{equation}
    \tau = -\frac{1}{a_0}\ln u_1,
    \label{eqn:tau}
\end{equation}
where $a_0$ is the sum of all propensities and $u_1\in (0,1)$ is a uniformly distributed random number.

Once the time until the next reaction is calculated, we draw another random number, $u_2$, to determine which reaction takes place. We select reaction $j\in\{1,\dots,4\}$ such that $\sum_{i=1}^{j-1}a_{i}<u_2a_0\leq \sum_{i=1}^{j}a_i$, with the convention that $\sum_{i=1}^{j-1} a_i=0$ for $j=1$. We then update the number of molecules of each species according to which reaction is selected.

For the BTS with division, we need not include effective dilution rates, given that we are considering division explicitly; thus, our propensities are as follows:
\begin{enumerate}
    \item $a_1 = \alpha_AH(\lambda_BV - N_B)$,
    \item $a_2 = \alpha_BH(\lambda_AV - N_A)$,
    \item $a_3 = \beta_AN_A$, and
    \item $a_4 = \beta_BN_B$.
\end{enumerate}
Again, the Heaviside functions must be evaluated at the beginning of each iteration; however, since the threshold values are proportional to the volume $V = V_0e^{\delta t}$, we must also calculate $V$ at the beginning of each iteration. 

We then calculate the time until the next reaction from Eq.~\eqref{eqn:tau}, but must ensure that the reaction occurs within the cell cycle. If $t$ is the cell cycle time and $t+\tau<T$, with $T$ the cell cycle duration, then we proceed as usual, determining which reaction takes place and updating the number of molecules accordingly. If, however, $t + \tau \geq T$, then the system is only advanced to $t=T$ and division occurs instead. At division, we partition the protein counts according to a binomial distribution with probability $1/2$ -- instead of simply dividing by two as in the deterministic case -- in order to maintain integer numbers of proteins. Note that for simplicity we have assumed that a timer mechanism is controlling division; more realistic cell-size homeostasis mechanisms, such as an adder or a sizer, could be incorporated into the simulation, as was done in \cite{jia2021cell}.

In calculating the levels of disagreement between the standard BTS and the BTS with division, simulations were run for 15 full cycles to allow both models to reach a steady state or cyclostationary state, respectively. At that point, the ``winning" protein is identified; if the winning proteins differ between the two models, the simulation is recorded as a disagreement.

\begin{figure}[H]
    \centering
    \includegraphics[width=0.5\linewidth]{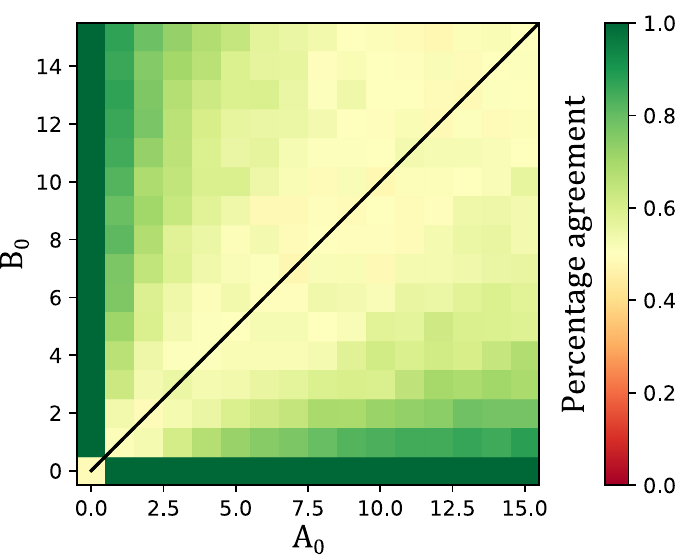}
    \caption{Absence of a deterministic ROD in a symmetric system with $\alpha_A=15=\alpha_B$, $\lambda_A=2=\lambda_B$, and $\beta_A = 0 = \beta_B$. Stochastic simulation may still result in disagreement due to noise pushing one trajectory over the separatrix and not the other.}
    \label{fig:symmetric_heatmap}
\end{figure}

As discussed in the Main Text, stochastic fluctuations alone can cause steady states to disagree between simulations of the standard BTS and the BTS with division, especially for initial conditions near the separatrix. Fig.~\ref{fig:symmetric_heatmap} demonstrates that even for parameter sets where there is no deterministic disagreement, approximately 50$\%$ of stochastic simulations initiated close to the separatrix will reach disagreeing steady states, giving a baseline for solely noise-induced disagreement. 

\section{Parameter sets}

Unless otherwise specified, all figures and analyses in the parameter regime where $\alpha_A\lambda_B>\alpha_B\lambda_A$ use the parameter set $\alpha_A=150$ [molecules/hour], $\alpha_B=15$ [molecules/hour], $\lambda_A=30\text{ [molecules]}=\lambda_B$, and $\beta_A=0 \text{ [molecules/hour]}=\beta_B$. In the regime where $\alpha_A\lambda_B<\alpha_B\lambda_A$, we instead take $\alpha_A=15$ [molecules/hour] and $\alpha_B=150$ [molecules/hour], again with $\lambda_A=30\text{ [molecules]}=\lambda_B=30$ and $\beta_A=0\text{ [molecules/hour]}=\beta_B$. Finally, $T=27.5$ hours throughout.


\end{document}